\documentclass[twocolumn, twocolappendix, usenatbib, useAMS]{aastex631}
\usepackage{fleqn, multirow, graphicx, color, enumitem}
\usepackage{array, mathtools, subfigure, esint,CJK}
\usepackage{soul, hyperref}
\usepackage{tikz}
\usetikzlibrary{arrows,shapes,chains}
\newcommand{\uat}[2]{\href{http://astrothesaurus.org/uat/#2}{#1 (#2)}}

\def\mpch{\,{h^{-1} {\rm Mpc}}}          \def\hmpc{\,{h {\rm Mpc}^{-1}}}

\makeatletter
\@for\next:={int,iint,iiint,iiiint,dotsint,oint,oiint,sqint,sqiint,
	ointctrclockwise,ointclockwise,varointclockwise,varointctrclockwise,
	fint,varoiint,landupint,landdownint}\do{%
	\expandafter\edef\csname\next\endcsname{%
		\noexpand\DOTSI
		\expandafter\noexpand\csname\next op\endcsname
		\noexpand\ilimits@
	}%
}
\makeatother
\shorttitle{Identifying subhalos using CWTHF}
\shortauthors{Li, Wang \& He}
\begin{document}
\begin{CJK*}{UTF8}{gbsn}
\title{CWTHF: Subhalo Identification with Continuous Wavelet Transform}

\author[0009-0003-1625-8647]{Minxing Li (李敏行)}
\affiliation{Center for Theoretical Physics, College of Physics, Jilin University, Changchun 130012, China.}

\author[0000-0003-4064-417X]{Yun Wang (王云)}
\affiliation{Key Laboratory of Material Simulation Methods \& Software of Ministry of Education, \\ College of Physics, Jilin University, Changchun 130012, China.}

\author[0000-0001-7767-6154]{Ping He (何平)}
\affiliation{Center for Theoretical Physics, College of Physics, Jilin University, Changchun 130012, China.}
\affiliation{Center for High Energy Physics, Peking University, Beijing 100871, China.}

\correspondingauthor{Ping He}
\email{hep@jlu.edu.cn}

\begin{abstract}
With advances in cosmology and computer science, cosmological simulations now resolve structures in increasingly fine detail. As key tracers of hierarchical structure formation, subhalos are among the most important objects within these simulations. In our previous work, we established that the continuous wavelet transform (CWT) can effectively extract clustering information and serve as a robust halo finder. Here, we extend the CWT framework to subhalo identification by adapting the Continuous Wavelet Transform Halo Finder (CWTHF) code. This extension extends the unbinding procedure, which enables the reliable identification of gravitationally bound substructures. The algorithm identifies density peaks within known halos or subhalos and segments the surrounding volume accordingly. Once a new subhalo is registered, its position is recorded to prevent duplicate detection. We validate our approach using the TNG50-2 and TNG100-1 simulations, as well as a single Friends-of-Friends (FOF) halo, by comparing the resulting CWT catalog against the reference SUBFIND catalog. Because the method inherits the original computational framework, our subhalo finder maintains a favorable linear time complexity of $\mathcal{O}(N)$.
\end{abstract}


\section{Introduction }
\label{sec:intro}

In the standard $\Lambda$CDM cosmological model, dark matter is composed of cold nonbaryonic particles (hereafter referred to as CDM), and structure formation proceeds in a bottom-up manner: low-mass halos form earlier and subsequently grow into more massive ones through merging and accretion. The assembly of a massive halo does not erase all information about its progenitors. Despite efficient relaxation mechanisms such as tidal stripping and dynamical friction, many substructures survive as self-bound entities known as subhalos. First proposed by \citet{White1978}, this scenario has been strongly supported by a wide range of observations. Evidence spans from the mere existence of massive galaxy clusters to the detailed substructure revealed by gravitational lensing \citep[e.g.,][]{Vegetti2010, Li2014}. 

While the detection of dark matter subhalos must rely on indirect, luminous, or dynamical tracers, cosmological simulations provide an ideal laboratory where subhalos and their abundance can be directly studied \citep{Ghigna1998, Tormen1998, Klypin1999}. In these simulations, the collective behavior of a discrete set of macroparticles mimics the dynamics of a continuous dark matter fluid. These particles form a pure $N$-body system, whose phase-space coordinates are evolved self-consistently through gravitational interaction.

Driven by rapid advances in observational facilities and computational power, cosmology has entered an era of precision, in which the challenge lies in extracting subtle, small-scale signals -- previously lost in the noise -- from the latest observational datasets. Given their small-scale nature and close connection to galaxies, subhalos have become indispensable probes of cosmic structure. They serve to calibrate observational measurements \citep{Gao2004, Guo2010, Contreras2021}, trace galaxy formation and evolution \citep{Otaki2023, Nadler2024, Sifn2024, Chandro2025}, quantify baryonic and tidal effects \citep{Heinze2024, Erkal2015, Du2025, Wang2025}, and even to diagnose the influence of the cosmic web \citep{Hunde2025}. Subhalo counts and properties can probe primordial curvature perturbations \citep{Ando2022} and also enable decisive tests of dark matter physics ranging from annihilating or decaying candidates \citep{Bringmann2009, Stucker2022} to self-interacting dark matter \citep{Dutra2025, Zeng2025, Zhang2025}.

The identification of substructures, however, requires the assignment of particles to specific subhalos when generating mock catalogs in simulations. The nonlinear nature of structure formation precludes a universally accepted definition for a dark matter halo, even less so for a subhalo \citep{Knebe2011, Onions2012, Knebe2013}. Embedded in a dynamically complex environment and subject to continuous tidal stripping, subhalos lack a distinct gravitational boundary. Consequently, delineating their extent is far more challenging than for isolated halos, as they are gravitationally inseparable from their host \citep{Mansfield2024}. Consequently, these definitional ambiguities have driven the proliferation of diverse subhalo-finding algorithms, each employing its own operational definition.

Halo finders that are capable of identifying substructures can be classified as subhalo finders. These methods are conventionally grouped into three categories. The first is the configuration-space finders which are defined by their use of particle positions to identify overdense regions as candidates for gravitationally self-bound substructures. The configuration-space approach includes HFOF \citep{Klypin1999}, the most widely used algorithm SUBFIND \citep{Springel2001} as well as subsequent developments such as ADAPTAHOP \citep{Aubert2004} and AHF \citep{Knollmann2009}. Although many incorporate velocity information to verify self-binding, these methods primarily operate in configuration space and thus remain fundamentally distinct from true phase-space finders.

The second is the phase-space finders, which incorporate particle velocities, either during peak searching \citep[e.g., VELOCIRAPTOR,][]{Elahi2019} or in the particle linking process \citep[e.g., ROCKSTAR,][]{Behroozi2013}. This allows them to distinguish between closely spaced subhalos that often merge in configuration-space \citep{Behroozi2015}. While this advantage is significant, it comes with a trade-off: the strict phase-space criteria can fail to identify tidally distorted subhalos as bound entities \citep{Diemer2024}.

To further trace and identify such subhalos, researchers often turn to their assembly history, leading to the last category: history space finders. Notable examples include SURV \citep{Giocoli2010}, HBT \citep{Han2012}, its later extension HBT+ \citep{Han2018}, and SYMFIND \citep{Mansfield2024}. The introduction of an additional time dimension successfully addressed these issues, albeit at the cost of significantly increased computational resources. Unlike methods that rely on a single snapshot, history-space finders must process multiple time-ordered snapshots. Moreover, their forward-tracking approach in time carries the risk of error amplification as the simulation progresses.

The application of wavelet transforms to the identification of cosmological structures began in the 1990s and has since become an established method in the field \citep{Slezak1990, Bijaoui1992, Slezak1993, Romeo2008}. This approach generally falls into two distinct yet functionally similar categories: the discrete wavelet transform (DWT) and the continuous wavelet transform (CWT). Demonstrating its versatility, the technique has been successfully applied to a wide range of data, from galaxy clusters \citep{Slezak1994, Girardi1997, Bardelli1998, Nikogossyan1999, Flin2006} and the Hercules moving group \citep{Liang2023} to narrow-band images of planetary nebulae \citep{Cuisinier2005} and combined angular-redshift samples of cluster galaxies \citep{Pagliaro1999}. In these diverse contexts, the wavelet transform proves to be a powerful tool for resolving substructures.

As demonstrated by \citet{Seymour2002}, the DWT can successfully identify subhalos within an individual dark matter halo. For the CWT, a comparative study with SUBFIND has shown that it is capable of detecting subhalos overlooked by this widely used finder \citep{Schwinn2018}. These successes highlight the exceptional capability of multiscale analysis in extracting information from the cosmic density field, effectively serving as a ``mathematical microscope."

In previous works, we verified that the CWT method provides a wavelet-based definition of dark matter halos and can function as a self-contained halo finder \citep{Li2024, Li2025}. Building on this foundation, the present study extends the same wavelet framework to subhalo identification and integrates this new capability into the Continuous Wavelet Transform Halo Finder (CWTHF) code. Adhering to the core CWT paradigm, we detect subhalos by identifying local maxima in the four-dimensional wavelet domain -- three spatial dimensions plus one scale. The three-dimensional spatial CWT volume surrounding these maxima is then segmented at the corresponding scales, enabling the determination of both the positions and precise spatial extents of individual substructures \citep{Bijaoui1992, Slezak1993, Flin2006}. Despite the incorporation of the new subhalo module, the execution time of CWTHF has not increased; in fact, optimization during integration reduced the total run time by 20\%. Beyond its primary function of analyzing full simulation data, the algorithm also serves as a standalone subhalo finder capable of processing individual halos while maintaining $\mathcal{O}(N)$ time complexity.

This paper presents an extension of the CWTHF code for subhalo identification, demonstrating its capability to extract substructures from cosmological simulation data. We introduce the CWT halo hierarchy and compare it with the conventional FOF+SUBFIND approach. The performance of our method is evaluated through a direct comparison with SUBFIND on a representative halo. The paper is structured as follows: Section~\ref{sec:method} details the CWTHF subhalo identification methodology; Section~\ref{sec:tng} describes the IllustrisTNG (TNG) dataset and its FOF+SUBFIND group structure; Sections~\ref{sec:full} and \ref{sec:single} present the CWTHF results for the full simulation box and an individual halo, respectively; and Section~\ref{sec:concl} provides concluding remarks on the subhalo identification capabilities of CWTHF.

\begin{figure*}[htb]
\centerline{\includegraphics[width=0.975\textwidth]{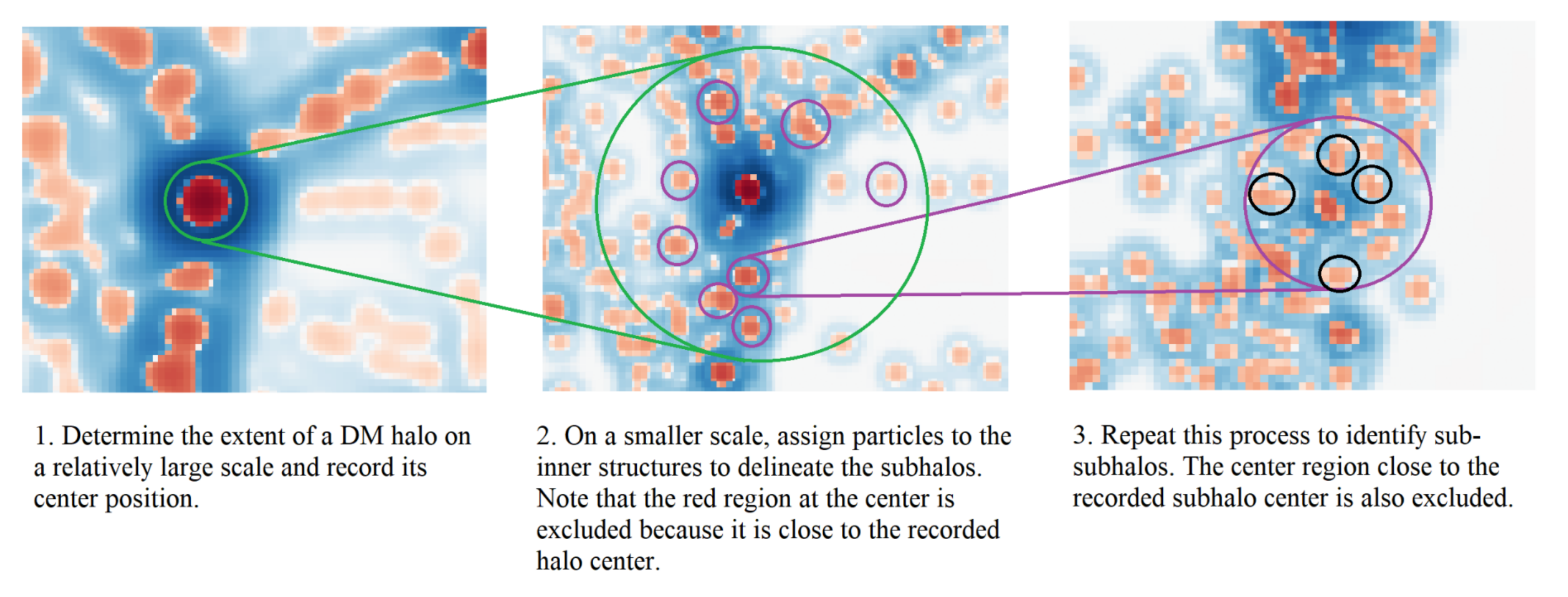}}
\caption{Visualization of the subhalo identification pipeline. Linked circles of the same color across adjacent panels mark the same physical region, while circles of the same color within a single panel represent structures (halos or subhalos) detected at that specific scale. Once a subhalo is confirmed, its center is recorded, and any subsequent maximum found within its exclusion zone is discarded. For example, the central red structure in Step~2 and the central orange structure in Step~3 were removed due to their proximity to a previously validated subhalo.}
\label{fig:visual_subhalo}
\end{figure*}

\begin{figure}[htb]
\centerline{\includegraphics[width=0.475\textwidth]{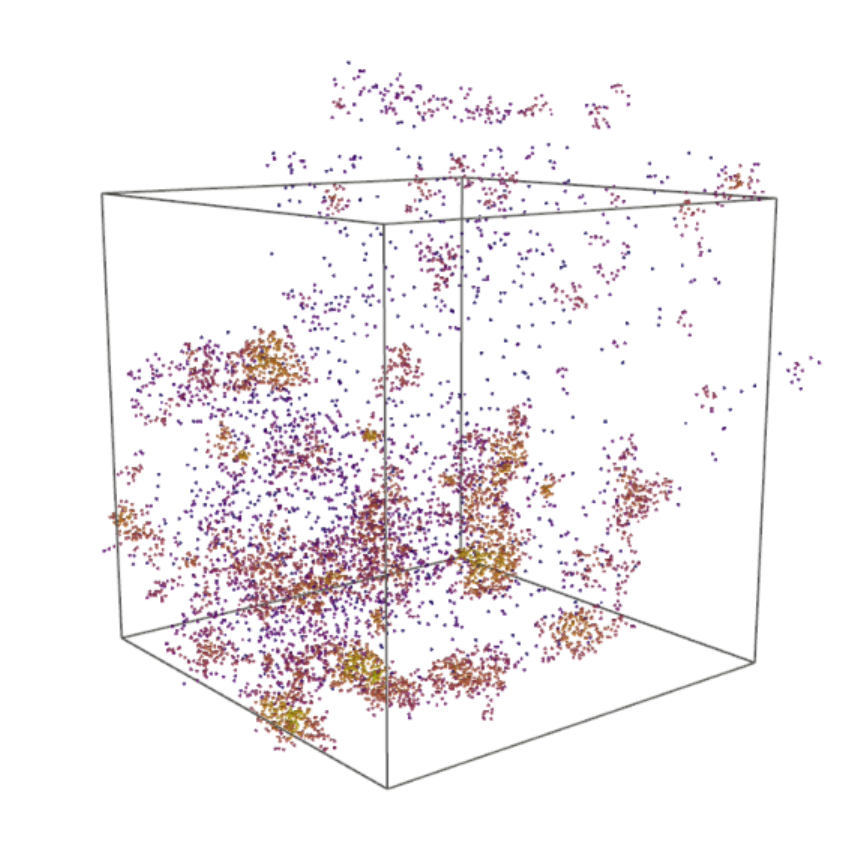}}
\caption{Scatter plot of a typical disrupted central subhalo. When maxima are not removed during identification, the nonorthogonality of the GDW generates a secondary peak at the same spatial location but on a smaller scale (higher $k_w$). This leads to the oversegmentation of the central subhalo from within, resulting in a hollow shell structure rather than a solid core.}
\label{fig:no_removal}
\end{figure}

\section{Methods}
\label{sec:method}

This section outlines the subhalo identification methodology employed in CWTHF. For comprehensive details on the underlying techniques, including the Gaussian-derived wavelet (GDW), CWT grid computation, spatial segmentation, peak correction, and the parallel domain decomposition approach. Readers are referred to \citet{Li2025}. Further specifics regarding the general CWT framework can be found in \citet{Wang2021, Wang2022a, Wang2022b}.

We describe the subhalo identification pipeline, with the full procedure visualized in Figure~\ref{fig:visual_subhalo}.
\begin{enumerate}
  \item Assign particles to a grid, compute the CWT, and identify local maxima. 
  \item Segment the CWT grid around these maxima, excluding those in close proximity to previously identified halos/subhalos.
  \item Assign particles to the segmented regions, where particles within the same segment are grouped as a candidate structure.
  \item Dispose of groups that either lack a cross-scale maximum or have a density below 4$\overline{\rho}$.
  \item Perform a self-boundness check on the remaining groups; confirmed halos/subhalos are added to the halo\_pos list.
  \item Determine and record the parent halo for each confirmed subhalo.
  \item Preserve gravitationally bound particles and assign a final halo ID.
\end{enumerate}

The subhalo identification process adheres to the established methodology, operating iteratively over a sequence of progressively smaller scale factors. Based on the user-defined resolution parameters, the code automatically computes the corresponding dimensionless scale parameter, $k_w$, for each iteration. Once the parameters are set, a geometric sequence of $\texttt{w\_resolution}$ elements is generated from a starting value of $\texttt{kw\_low}+l$ to an ending value of $\texttt{kw\_high}+l$, where $l=1.2$ is a parameter that has the effect of flattening the distribution of the entire sequence.

Since $k_w$ is inversely proportional to physical scale, a true sampling would require a sequence equally spaced in $1/k_w$. However, because the grid resolution is directly tied to $k_w$, such a $1/k_w$ sequence would result in excessively large intervals between iterations at small scales. Our chosen approach balances these two competing factors, producing a sequence that provides finer sampling at smaller $k_w$ and coarser sampling at larger $k_w$, thereby avoiding any large gaps\footnote{The CWTHF program is robust to the specific form of this sequence, provided this core sampling requirement is met.}.

After the $k_w$ sequence is determined, the grid resolution $N_g$ and the wavelet parameter \texttt{gdw\_resolution} for each $k_w$ are assigned as follows:
\begin{flalign}
    &N_g = {\rm Int}(\texttt{n\_ref}\times k_w + \texttt{n\_min}) \nonumber \\
    &\texttt{gdw\_resolution} = {\rm Int}(1.5/k_w+5). \nonumber 
\end{flalign}
Once these parameters are established, CWTHF proceeds with its first iteration. Particles are assigned to a regular grid using the cloud-in-cell (CIC) scheme, from which the CWT grid is computed. Local maxima in the CWT are then identified. At this initial stage, the halo list is empty, so no maxima are discarded based on proximity to existing structures. The amplitudes of the detected peaks are refined, two auxiliary arrays storing peak information are updated, and the CWT grid is segmented. This sequence completes one iteration, after which the procedure continues to the next $k_w$ value.

Beginning with the second iteration (i.e., for a larger $k_w$), the procedure follows the same steps as the first, with one addition: before locating CWT maxima, temporary IDs are assigned to all particles, and further halo/subhalo confirmation is performed. This adjustment is necessary because the cross‑scale verification at the $n$-th $k_w$ requires both the CWT maxima from the current iteration and the full CWT grid computed at the $(n+1)$-th $k_w$. All subsequent iterations follow this updated routine. When the algorithm reaches the final $k_w$ in the sequence, no finer-scale CWT grid is available. In this case, a zero-valued grid is supplied as the $(n+1)$-th grid. This iteration corresponds to the smallest resolvable scale; consequently, all detected maxima automatically satisfy the cross‑scale verification, as the resolution limit has been reached.

In the final step, an additional self-boundness check is applied to all identified halos and subhalos to verify whether any particles become unbound after the extraction of substructures. In rare cases, a halo may cease to be gravitationally bound once the potential contribution from its subhalos is removed, and such systems are consequently excluded from the final halo catalog.

As previously discussed, the wavelet method identifies structures solely based on their scales \citep{Li2024, Li2025}. This feature automatically filters out the influence of satellites when calculating the CWT grid, thereby preventing the identification of the host halo (and the central subhalo) from being affected by assembly bias.

However, our unique segmentation pipeline introduces two differences compared to SUBFIND. The first is the additional 30\% particle content in each central subhalo. To obtain a smooth boundary, CWTHF extends the spatial occupation of every structure as far outward as possible to the CWT valley. These particles in the outermost region are tightly bound to the structure and remain bound to it even under stricter escape velocity thresholds. Secondly, our CWT method assumes that any subhalo must be unimodal. This means that an outer density peak will be excluded if this peak has a similar scale -- that is, it corresponds to another peak in the CWT map -- to the central subhalo. The difference between the subhalo definitions adopted by the two methods automatically distorts their shape, consequently affecting the rank of the largest halos shown in Figure~\ref{fig:Top_subs}.

\subsection{Removal of the Repeated Structures}

Notably, the key difference between the current subhalo identification approach and the previous program is that the particles in the identified halos will not be removed, but are instead retained in the segmented grid to verify their subhalo membership. This approach, however, introduces a severe problem of repeated identification. As noted in our earlier work, the GDW we adopt is isotropic and has approximately compact support, making it nonorthogonal \citep{Daubechies1992}. Due to this nonorthogonality, a single structure can generate multiple local maxima across different scales. Such multiplicity severely disrupts the identification pipeline, since the CWTHF algorithm relies precisely on these maxima to locate halos and subhalos. An illustration of this issue is provided in Figure~\ref{fig:no_removal}.

To address this issue, we record the coordinates of every confirmed halo or subhalo and discard any newly detected maximum that lies within a fixed neighborhood of these stored positions. Since the iterations proceed from large to small scales, this procedure always preserves the first cross-scale maximum encountered at a given location -- corresponding to the largest structure identified there. A direct pairwise comparison between every detected maximum and all stored halo/subhalo positions would be computationally prohibitive: given $\sim$$10^7$ maxima and $\sim$$10^6$ confirmed objects, approximately $10^{13}$ distance checks would be required, which is infeasible in practice.

To address this, we map the recorded positions onto a separate grid of identical dimensions, specifically, a three-dimensional array of shape $N_g^3$, so that the maxima extracted from the CWT grid can be compared via a simple array lookup. Each halo/subhalo position is assigned to this grid using the nearest-grid-point (NGP) scheme and is then expanded by one cell in all directions, creating a $3\times3\times3$ exclusion zone centered on the corresponding grid point. Once this mask is constructed, each candidate maximum is checked with a single array access: if its grid index lies within any marked exclusion zone, the peak is discarded and excluded from subsequent segmentation.

The introduction of the maximum filter incurs negligible overhead in both runtime and memory consumption. The assignment and indexing operations are both $\mathcal{O}(N)$, amounting to at most $\sim 10^{7} + 10^{6}$ steps in total. Although the mask array has the same shape as the CWT grid, it is stored as a Boolean array, which occupies only one-quarter of the memory of a floating point array and is discarded immediately after the local maxima detection step. Thanks to this rapid build-and-discard approach, the overall runtime and memory footprint of the CWTHF pipeline remain effectively unchanged.

\subsection{Substructure Extraction and Hierarchy Buildup}

The correspondence between CWT maxima and dark matter halos was established in our earlier work. A similar relationship has been reported for subhalos extracted from individual host halos, although that study applied a threshold to the wavelet coefficients rather than adopting the peak finding scheme used here \citep{Schwinn2018}. To reduce contamination from the host halo background, those authors fitted the host mass distribution with a Navarro–Frenk–White (NFW) profile \citep{Navarro1996} and subtracted its contribution prior to performing the CWT. By contrast, our approach does not require such a background fitting and subtraction step.

As previously mentioned, subhalos are always embedded in complex dynamical environments, with the host halo posing a particular challenge. While fitting and removal strategies can mitigate this issue, they do not fully leverage the analytical capabilities of the CWT. In most cases, the background halo is substantially more massive and extended than the embedded subhalos. Although the CWT alone cannot directly disentangle substructures from such a dominant background \citep{Baluev2020, Cui2022}, multiscale analysis remains capable of detecting them across different, and even adjacent, scales \citep{Bijaoui1992, Flin2006}.

In this study, we extend the capability of the CWT for subhalo identification by incorporating an unbinding procedure that separates substructures from their background. Based on the principle that any physical structure produces a local maximum in the CWT grid, we assign each particle, even those already belonging to confirmed halos or subhalos, to the segmented region obtained in each iteration and provide it with a temporary group ID. Whenever a group meets the criteria of sufficient density, hosts a cross-scale maximum, and is self-bound, it is ``cut out" from its parent halo and assigned a new, permanent halo ID. After the particle membership of the subhalo is finalized, we record its parent group (i.e., the structure from which it was detached) before updating the halo catalog. The parent labeling is maintained consistently: ID 0 indicates a central subhalo rather than a hierarchically embedded subhalo, while a nonzero ID points to its immediate parent. In this manner, a hierarchical halo catalog is systematically constructed.

Regardless of how blended a halo and its subhalos may be, an isolated subhalo must manifest as a density peak in the dark matter density field. This assumption underpins virtually every subhalo finder, and our CWTHF is no exception. Through multiscale peak analysis, their positions and sizes are located, producing a series of local maxima in the 4D wavelet space. An additional self-boundness check not only smooths the sharp boundaries introduced by grid-based segmentation, but also significantly enhances the discriminating power of the CWT technique, enabling it to cleanly extract substructures from dense backgrounds.

\section{The Data}
\label{sec:tng}

In this study, we utilize two snapshots from the TNG project to test our program. The TNG project is among the most well-known and state-of-the-art cosmological simulations of galaxy formation \citep{Nelson2019a}. It is built on the moving-mesh code AREPO, in which the evolution of each dark matter particle and gas cell is governed by coupled gravity-magnetohydrodynamic equations. These equations are solved on a dynamically unstructured mesh using a second-order accurate Godunov-type scheme \citep{Springel2010}. To better reproduce the real universe, the simulation incorporates a comprehensive set of astrophysical processes when updating the particles and cells, including black hole formation and feedback, star formation and stellar feedback, metal enrichment and associated gas cooling, as well as supernovae.

The simulation evolves from $z=127$, incorporating the subgrid physics modules mentioned above and adopting the Planck cosmological parameters: $h=0.6774$, $\Omega_{\rm \Lambda } = 0.6911$, $\Omega_{\rm m}  = 0.3089$, $\Omega_{\rm b}=0.0486$, $n_s = 0.9667$, and $\sigma_8 = 0.8159$ \citep{Planck2016}. For its three flagship runs, TNG50, TNG100, and TNG300, the project provides 100 snapshots spanning $z=20$ to $z=0$. The names correspond to the side length of the simulation box in units of $\rm Mpc$. Each flagship run is accompanied by several lower-resolution versions that share the same physical settings but employ reduced numbers of particles and cells.

Among these simulation runs, we primarily utilize $z=0$ TNG50-2 for subhalo analysis. For single-halo tests and to probe the performance limits of our CWTHF method, we employ the full-resolution $z=0$ TNG100-1 run\footnote{The decision to use TNG100-1 instead of continuing with TNG50-1 for this test is motivated by its lower particle count. Loading an excessively large number of particles reduces the upper limit of $k_w$, which can cause the loss of small-scale halos/subhalos.}. TNG50-2 has a box size of $35\mpch$ and contains $1080^3$ dark matter particles, whereas TNG100-1 has a box size of $75\mpch$ with $1820^3$ particles. For the single-halo analysis, we select the most massive halo in TNG100-1, zoom into its region to resolve its finest structures, and directly compare them with the subhalos identified by SUBFIND. In the following analysis, only dark matter particles are considered.

\subsection{The Halo Hierarchy of the TNG Simulations}

The TNG snapshots include the halo and subhalo membership for each particle. To optimize storage, particles are arranged sequentially according to their group membership; consequently, individual particle IDs are not stored per halo/subhalo. These grouped particles correspond to FOF halos. The SUBFIND algorithm is then applied to each FOF halo to identify gravitationally bound substructures, thereby establishing a hierarchical FOF halo-SUBFIND subhalo catalog.

The construction of FOF halos follows a relatively straightforward linking procedure. Particles separated by less than $0.2l_{\rm aver}$, where $l_{\rm aver}$ denotes the average interparticle distance, are connected into groups. Groups containing fewer than a threshold number of particles are discarded, and the remaining groups are recorded as FOF halos \citep{Davis1985}. In contrast, the subsequent division of each FOF halo into gravitationally bound SUBFIND subhalos involves a more complex algorithm \citep{Springel2001}.

SUBFIND begins by estimating the local density for each particle at its position and sorting all particles by density. It then identifies local density peaks by searching, for each particle, the $N_{\rm ngb}$ nearest neighbors according to this density ranking and retaining the two neighbors with the highest density. Each particle either forms a new subgroup, is attached to an existing subgroup, or serves to link two existing subgroups when these two neighboring particles both have lower density, when one or both belong to the same existing subgroup and have higher density, or when they have higher density but belong to different subgroups, respectively.

\begin{figure*}[htb]
\centerline{\includegraphics[width=0.975\textwidth]{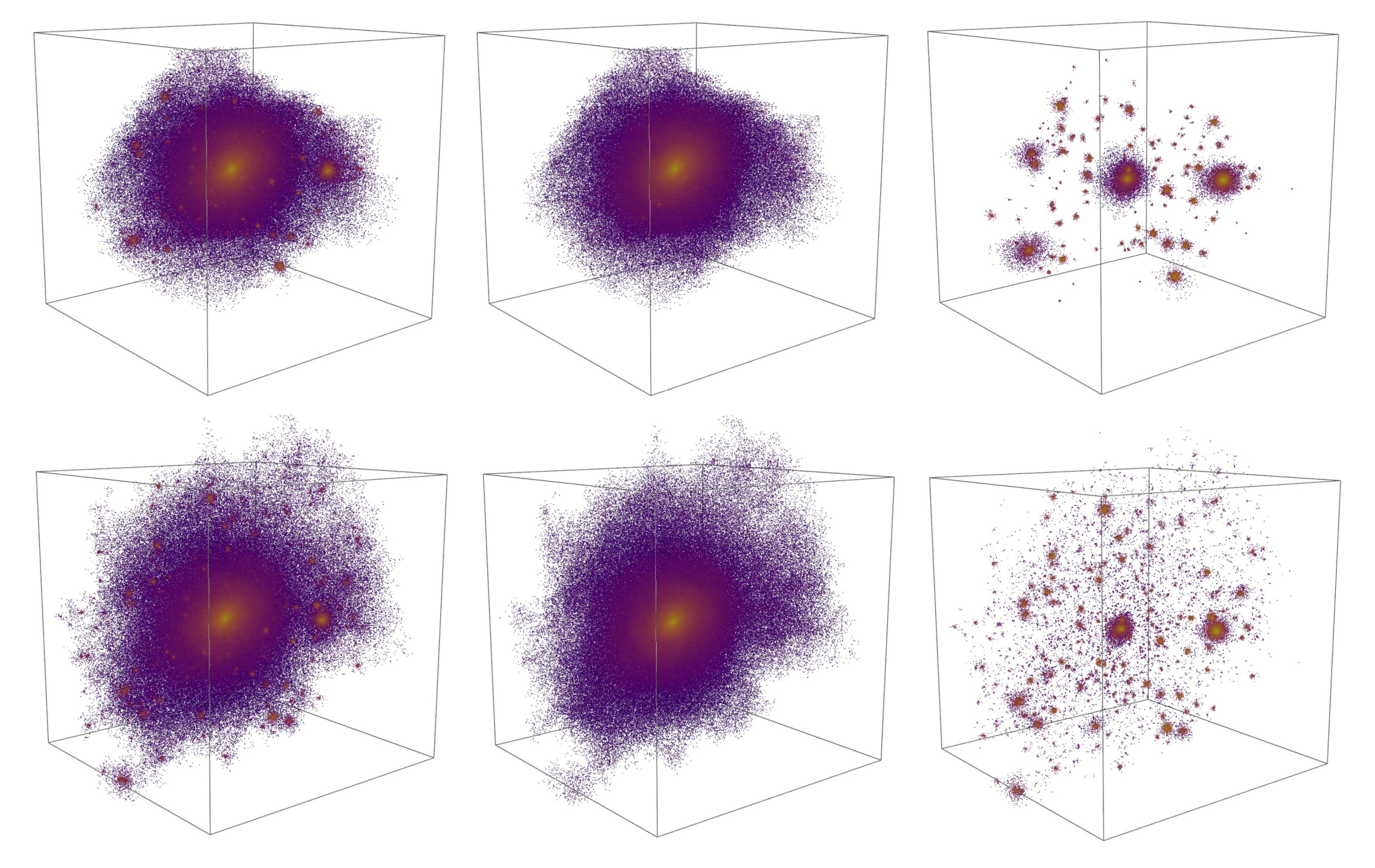}}
\caption{Scatterplot of the largest CWT halo (top panel) and its corresponding FOF halo (bottom panel) with their color encoding the local density. Each row, from left to right, shows the full halo, the central subhalo, and all remaining subhalos, respectively. The cube edge length is $2\mpch$. Notably, the CWT halo is more compact, excluding the outermost regions of the FOF halo. The significantly more concentrated distribution in the subhalo panels, together with the small high-density patches visible within the central subhalo, indicates a systematic trend: in full-snapshot analysis, CWTHF captures all major substructures but tends to leave the smallest subhalos merged with the central subhalo.}
\label{fig:CWT_SUBFIND_tot}
\end{figure*}

Since the merging of subgroups does not overwrite their original IDs, a particle may temporarily be associated with multiple subgroups. The algorithm first performs a self-boundness check on all candidate subgroups. Particle IDs are then updated from those of larger subgroups to the corresponding smaller ones, ensuring each ID reflects the smallest subgroup to which the particle belongs. After every particle is assigned a unique subgroup membership, a final self-boundness check is applied. The bound subgroups that survive this process form the final subhalo catalog.

To facilitate a direct comparison with the TNG catalog, we first preprocess the simulation data. In the TNG snapshots, FOF halos are constructed from all particle types (dark matter, gas cells, etc.). In the TNG50-2 run, 364,242 of the 1,316,275 FOF halos contain fewer than 20 dark matter particles. Moreover, while every object in our CWT catalog is self-bound, FOF halos are not inherently guaranteed to be. We therefore match our detections against the SUBFIND subhalo catalog, which applies a self-boundness check to each FOF halo, consistent with the criterion used in our method. For consistency, we also exclude 11,493 SUBFIND subhalos (out of 859,077) that contain fewer than 20 dark matter particles.

To align with the structure of the original TNG catalog, we associate each subhalo within an FOF halo to its most massive central subhalo. The only difference in our postprocessing is that we additionally discard particles not belonging to any bound group. The resulting halo is defined as the union of all retained subhalos; these postprocessed halos are referred to as FOF halos throughout the rest of this paper. For the CWT catalog, the central subhalo is identified directly from the particle data, and the full CWT halo is reconstructed using the hierarchical information generated by CWTHF.

Throughout this paper, all subhalo statistics (SHMFs, SHPS, etc.) include the central subhalo, consistent with our convention that the term ``subhalos" refers to all gravitationally bound substructures within an FOF group.
\section{Identification of the full snapshot}
\label{sec:full}

This section provides a detailed comparison between the CWT halo hierarchy and the traditional FOF-SUBFIND hierarchy in the full TNG snapshots. Our analysis focuses on two different cases: a sufficient run \texttt{0.1-5.5\_d4\_r400\_w35} toward the smallest halos in TNG50-2, and \texttt{0.1-7.0\_d4\_r400\_w55} with slightly limited resolution in TNG100-1. Taking the former as an example, its parameters are set as follows: \texttt{kw\_low}=0.1, \texttt{kw\_high}=5.5, \texttt{dens\_th}=4, \texttt{n\_ref=400}, and \texttt{w\_resolution}=35. These parameters follow the same definitions as in \citep{Li2025}. 

\begin{figure}[htb]
\centerline{\includegraphics[width=0.475\textwidth]{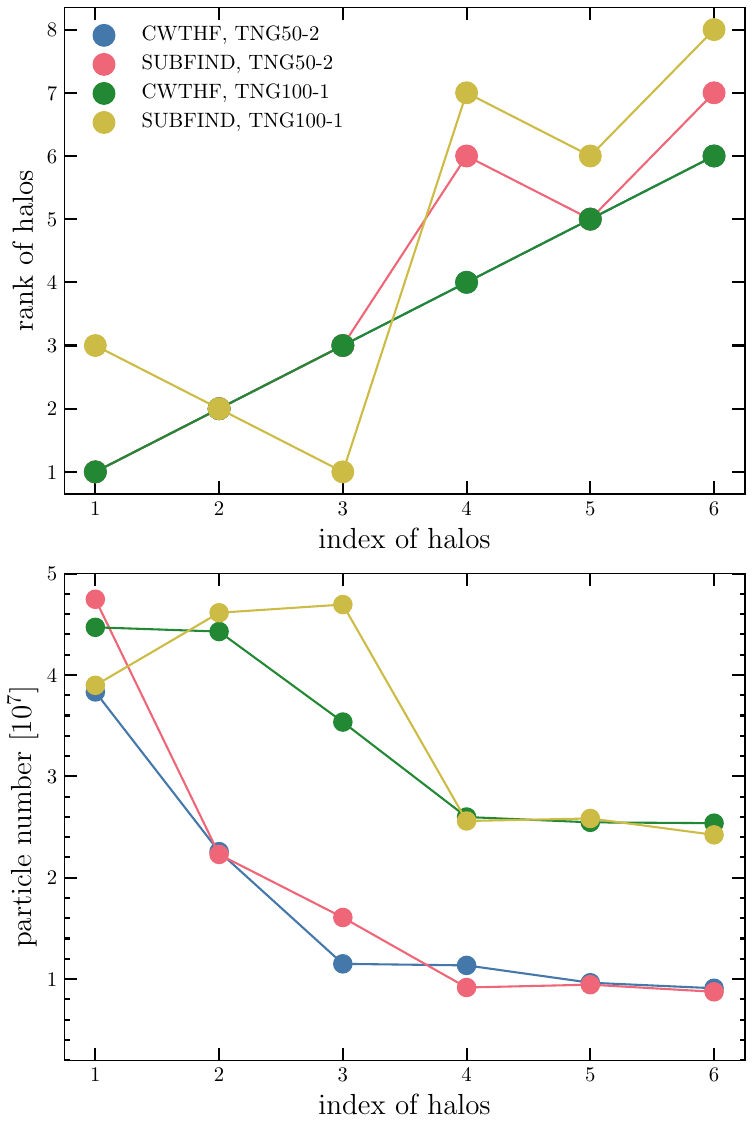}}
\caption{The rank (top) and the mass (bottom) of the six largest subhalos in the CWT catalogs, alongside their counterparts in the SUBFIND catalogs. The difference between the two finders is so minute that each corresponding pair of curves lies in close proximity.}
\label{fig:Top_subs}
\end{figure}

In general, CWTHF identifies roughly 30\% more bound particles residing in subhalos than SUBFIND does. Specifically, the CWTHF catalogs contain 841,521,628 and 4,259,489,642 particles that reside in 933,091 and 3,384,281 subhalos for TNG50-2 and TNG100-1, respectively, whereas the corresponding SUBFIND catalogs contain 673,835,987 and 3,138,578,077 particles that reside in 847,584 and 4,307,873 subhalos, respectively. Figure~\ref{fig:CWT_SUBFIND_tot} displays the largest CWT halo and its subhalo decomposition obtained with CWTHF in TNG50-2. The corresponding FOF halo and its SUBFIND subhalos are shown in the bottom panel for comparison. In these scatter plots, each particle is colored according to its local density estimated using a Gaussian kernel, and a logarithmic color map is used to highlight internal substructures. The two halos, along with their several most massive subhalos, exhibit similar positions, shapes, and particle content across the two catalogs and are therefore considered mutual counterparts.

Both in the CWT and SUBFIND catalogs, the central subhalo retains the overall shape of its parent halo, despite the fundamentally opposite construction strategies: CWTHF assembles subhalos upward into a halo, while SUBFIND splits an existing FOF halo downward into subhalos. The fixed edge length of the background cube indicates that the CWT halo and its central subhalo are smaller and notably more compact than their FOF/SUBFIND counterparts. This trend aligns with the findings of our earlier work, confirming that the halo identification capability of our method remains robust. Specifically, the CWT halo consists of 41,106,856 particles, with 38,325,117 located in the central subhalo and the remaining 2,781,739 distributed among other subhalos. In comparison, the corresponding FOF halo contains 52,682,598 particles, of which 47,469,006 belong to the central subhalo and 5,213,592 to the remaining subhalos.

The difference in overall halo size and central subhalo size arises mainly because the CWT halo definition deliberately excludes diffuse outer structures, treating them instead as separate, isolated halos. However, a more pronounced discrepancy is evident in the right panel, where CWTHF identifies far fewer satellite subhalos than SUBFIND does. While the SUBFIND panel is densely populated with dots representing numerous small subhalos, the CWT catalog displays only the most prominent ones. Consequently, the CWT halo comprises merely 154 subhalos, a sharp reduction compared to the over 10,000 subhalos in the corresponding SUBFIND catalog.

It is evident that mere exclusion of outer structures cannot explain the discrepancy of nearly 2 orders of magnitude in subhalo counts. Furthermore, the presence of numerous minor density peaks within the CWT central subhalo provides additional evidence that small, bound substructures remain coupled to the central object. The primary reason for this loss of detection is, in fact, an insufficient scale range in the wavelet analysis. As noted earlier, subhalos are often heavily tidally stripped, losing much of their outer dark matter content. Their compact morphology, combined with the complex background density of their host halo, requires a much finer spatial segmentation, that is, a higher value of \texttt{kw\_high}. A detailed analysis of this effect is presented in Section~\ref{sec:single}, which confirms that a sufficiently high \texttt{kw\_high} can significantly mitigate the issue.

To evaluate the consistency between the two methods, we select the top six CWT subhalos and their SUBFIND counterparts from two different simulation runs. Their ranks and masses are shown in Figure~\ref{fig:Top_subs}. The distributions of the most massive subhalos differ noticeably between the runs, reflecting the combined effects of differing resolution and box size. In TNG50-2, the largest subhalo contains roughly twice as many particles as the second-ranked one. In contrast, the decline in subhalo mass is much more gradual in TNG100-1, where the top three subhalos differ by only about 25\% in mass.

Free from contamination by unbound particles and spurious linkages due to the ``linking bridge" problem, the subhalo ranking is significantly more stable between the two catalogs than in the halo case, showing only minimal ($\leq3$) differences. In terms of subhalo mass, the relative difference drops to at most 30\% for the largest subhalos and remains below 10\% on average. This consistent pattern holds for both paired configurations, despite variations in the simulation parameters and CWTHF settings.

\begin{figure}[t!]
\centerline{\includegraphics[width=0.475\textwidth]{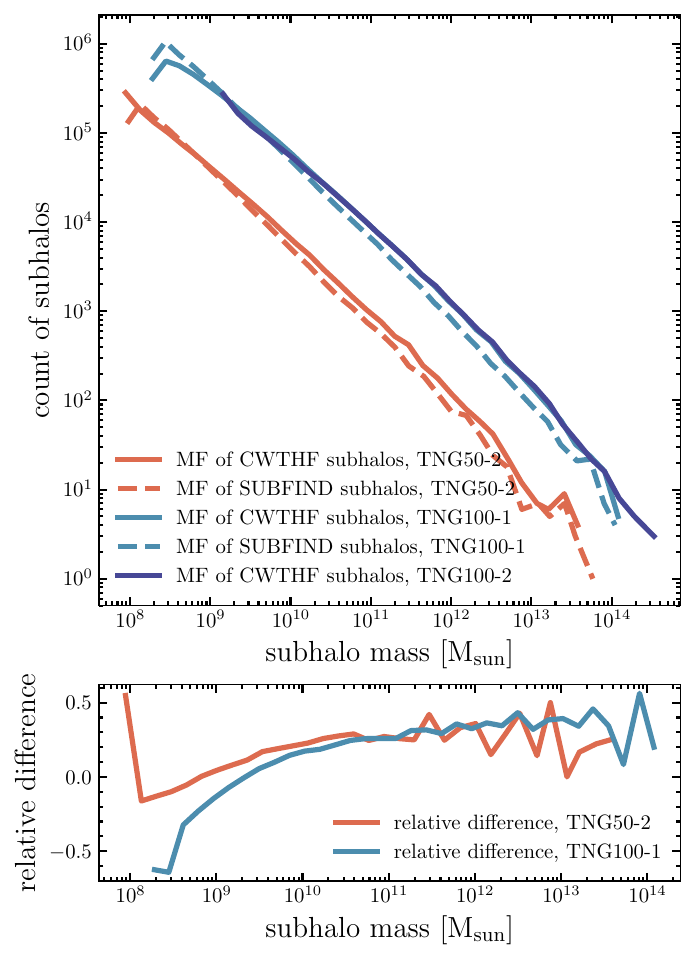}}
\caption{The SHMFs of SUBFIND and CWT subhalos (top) and their relative differences within each run (bottom). The red and blue lines represent the TNG50-2 and the TNG100-1, respectively, while the purple line represents the TNG100-2 for a test of the resolution effect of the simulation. The dashed and solid lines indicate the SUBFIND and CWTHF methods, respectively. The relative difference is defined as $(\rm N_{CWTHF}-\rm N_{SUBFIND})/\rm N_{CWTHF}$.}
\label{fig:HMF_tot}
\end{figure}

To further examine the mass distribution of CWT subhalos, we analyze the subhalo mass functions (SHMFs), shown in Figure~\ref{fig:HMF_tot}. The mass functions indicate that the value of \texttt{kw\_high} used for TNG50-2 is higher than in our earlier halo-finding runs, enabling the detection of more low-mass subhalos. This increase in detected substructures raises the amplitude of the mass function, resulting in an excess of approximately 50\% at the lowest mass end. In TNG100-1, the shape of the mass function and its relative difference show an approximate 50\% deficit, a behavior that aligns closely with, and is nearly identical to, the halo mass function presented in our previous study.

As the particle number exceeds 300, the relative differences become less pronounced. It is evident that in both runs, the values rise gradually from zero to over 30\%, where they stabilize until the particle number surpasses $10^5$. In the TNG50-2 snapshot, the insufficient number of large subhalos leads to fluctuations in the relative difference. Notably, this numerical instability only becomes relevant for TNG100-1 when the particle number exceeds $3\times10^6$. This highlights a rather expected distinction: with more massive macro DM particles and a larger box length, TNG100-1, as anticipated, contains more large subhalos, resulting in a mass function that is both higher in amplitude and smoother across all mass bins.

\begin{figure}[t!]
\centerline{\includegraphics[width=0.475\textwidth]{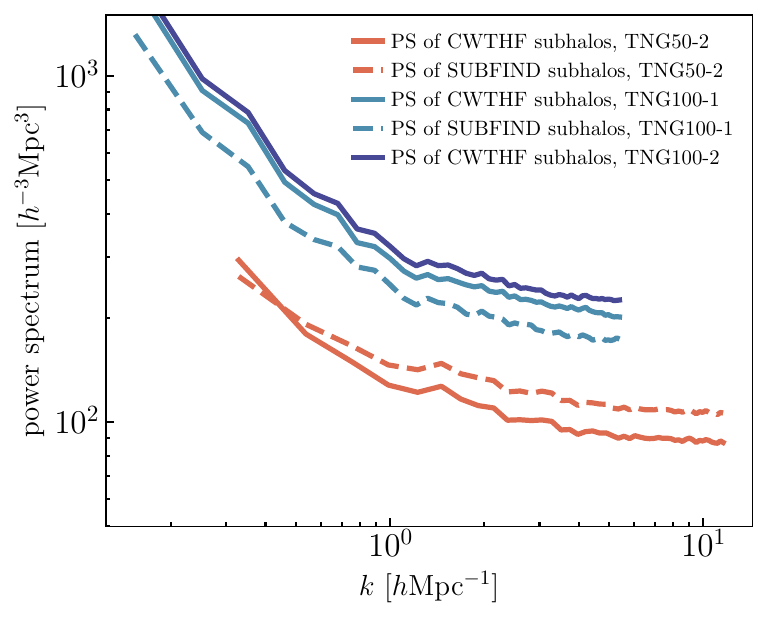}}
\caption{The SHPS of SUBFIND and CWT subhalos. The line styles and color coding follow the same scheme as in Figure~\ref{fig:HMF_tot}. Within each pair, the curves are highly similar, such that every fluctuation occurs in the same direction.}
\label{fig:PS_tot}
\end{figure}

Figure~\ref{fig:PS_tot} displays the subhalo power spectra (SHPS) for four catalogs from different simulations. These spectra were obtained by assigning all subhalos to a $256^3$ grid using the CIC scheme, weighting each by its mass. Since the TNG100-1 catalogs contain a larger number of subhalos, their power spectra consistently show higher amplitudes on all scales.

All SUBFIND subhalos are inherently bound structures, which leads to a closer match between the two corresponding catalogs than in the halo case. Although their power spectra are not identical, they follow very similar evolutionary trends. After applying appropriate scaling adjustments, their power spectra can be brought into close alignment. This consistency reinforces the reliability of the spatial distributions identified by the CWT algorithm. In both simulation runs, the SHPS of the CWT catalog exhibits a higher amplitude than SUBFIND on the largest scales. However, as $k$ increases, the SHPS of TNG50-2 shows a declining trend, eventually falling below that of SUBFIND once $k$ exceeds $0.5\hmpc$.

To test the resolution effect, we applied the same CWTHF parameters (\texttt{0.1-7.0\_d4\_r400\_w55}) to TNG100-2. As shown by the purple curves in Figure~\ref{fig:HMF_tot} and Figure~\ref{fig:PS_tot}, the simulation resolution does not affect the SHMF; it only raises the SHPS amplitude slightly, while its overall trend remains almost unchanged. The consistency between TNG100-1 and TNG100-2 demonstrates the resolution stability of the CWTHF within the same simulation series.

\begin{figure}[t!]
\centerline{\includegraphics[width=0.475\textwidth]{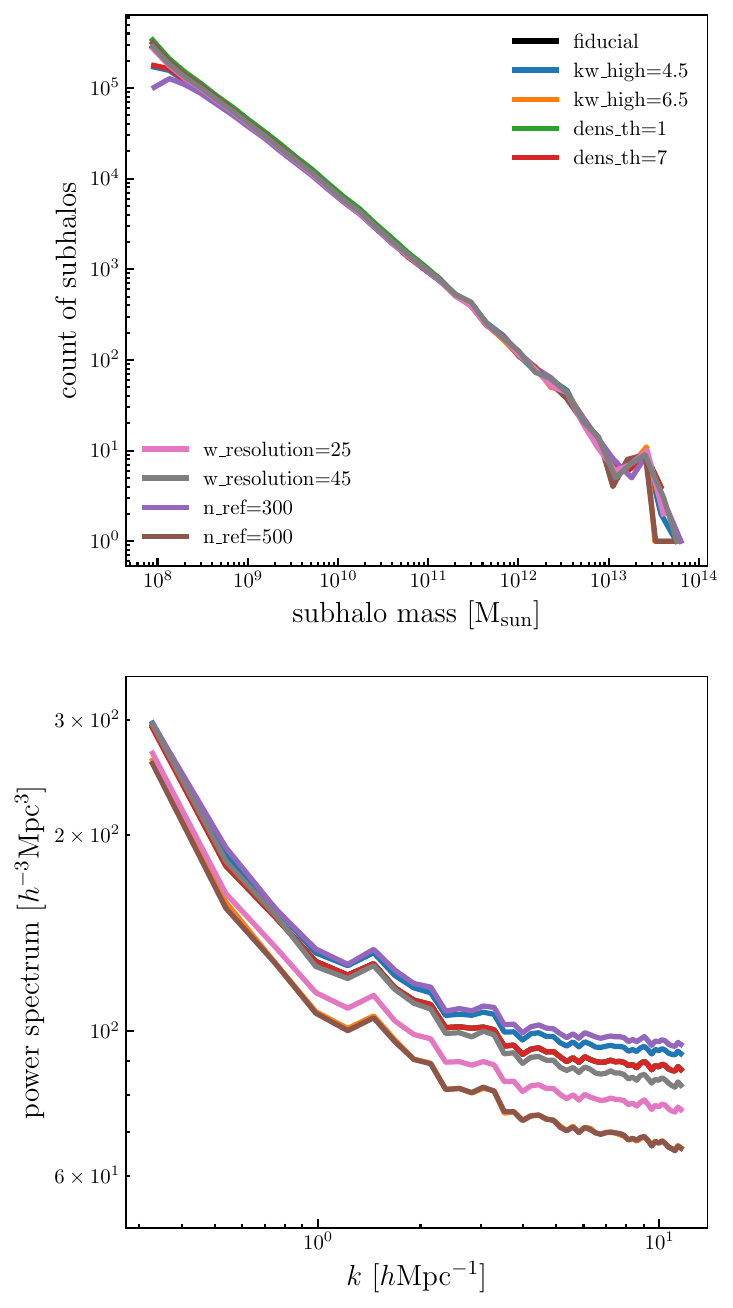}}
\caption{The SHMFs (top) and SHPS (bottom) of CWT subhalos in TNG50-2, each varying one parameter relative to \texttt{0.1-5.5\_d4\_r400\_w35}. Both of them show a remarkably high degree of consistency across all parameter changes.}
\label{fig:para_eff}
\end{figure}

In contrast to TNG100-1, where the SHPS of the CWT catalog remains consistently higher than that of SUBFIND, TNG50-2 exhibits a crossover between the two. This does not reflect a physical difference in the CWT catalog, but rather a parameter effect arising from the limited range of $k_w$. Figure~\ref{fig:para_eff} illustrates these parameter effects for CWTHF in TNG50-2. It is evident that increasing the number of small subhalos, whether by raising \texttt{kw\_high} or \texttt{n\_ref}, drives the SHPS further downward. Notably, setting \texttt{kw\_high=6.5} or \texttt{n\_ref=500} results in a coincidental overlap of the two power spectra. Meanwhile, \texttt{dens\_th} has only a minor influence on the SHPS, as indicated by the yellow and green lines obscuring the fiducial result. Additionally, increasing \texttt{w\_resolution} has almost no impact on the SHPS, since no new subhalos are identified; however, an insufficient scale resolution causes the SHPS to bend to lower amplitudes.

Regarding the SHMFs, the parameters affect only the low-particle-number regime, specifically where counts fall below $10^2$. The resulting trends are clear: higher resolution or looser thresholds elevate the SHMF, while lower resolution or stricter thresholds suppress it. As the subhalo mass increases, all SHMFs converge to identical values across every mass bin. Even at the high-particle-number end, any apparent difference is merely reflected in a subhalo count variation of at most a few objects.

This paper does not present a new, comprehensive performance benchmark, as the computational framework directly follows that of \citet{Li2025}. Through implementation-level optimizations, we achieved an approximately 20\% reduction in execution time. For reference, the run \texttt{0.1-5.5\_d4\_r400\_w35} on TNG50-2 was completed in 18,000 seconds, and \texttt{0.1-7.0\_d4\_r400\_w55} on TNG100-1 in 100,000 s, both using 20 MPI processes.

\section{Identification of a Single Halo}
\label{sec:single}

In Section~\ref{sec:full}, we examine the decomposition of a DM halo using CWTHF in the full TNG snapshot. Compared to SUBFIND, the separation of satellite subhalos from the central subhalo in CWTHF is far from complete. This is evident not only in the significantly smaller number of subhalos identified, roughly 2 orders of magnitude fewer, but also in the scatter plot, where some small subhalos remain coupled to the central subhalo.

\begin{figure}[t!]
\centerline{\includegraphics[width=0.475\textwidth]{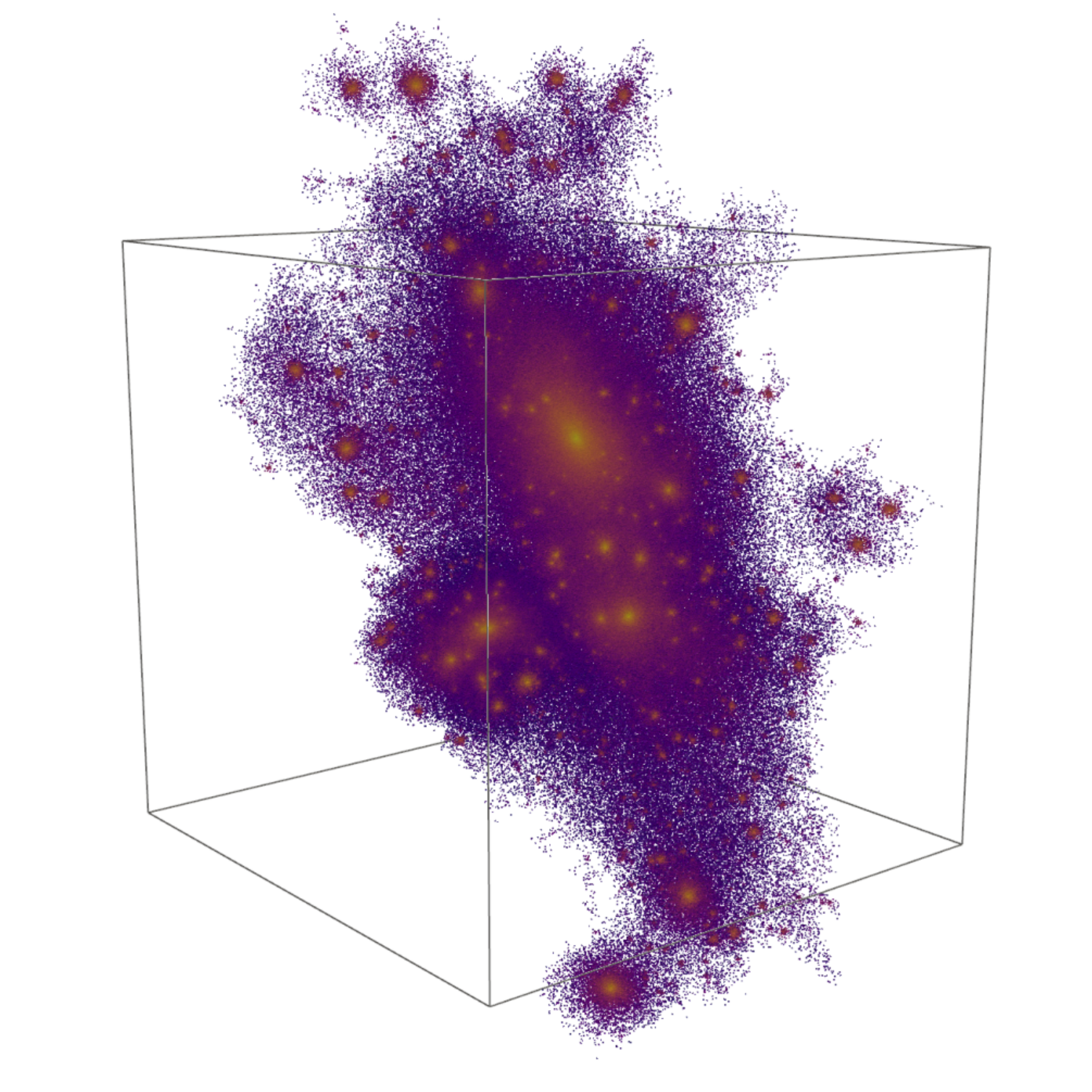}}
\caption{The scatterplot of the largest FOF halo in TNG100-1. The edge length of the cube is $3\mpch$.}
\label{fig:largest_FOF}
\end{figure}

The substantial discrepancy observed between different subhalo finders cannot be attributed to the halo definition alone, which suggests that the CWT halo is more compact and contains less of the outer diffuse material. To overcome the limitation imposed by the initial scale range selection and further explore the potential of our CWT method, we apply a zoom-in identification to the largest halo in TNG100-1. This halo, with a scale of approximately 3$\mpch$, contains 66,511,903 particles; its detailed morphology is presented in Figure~\ref{fig:largest_FOF}. Additionally, we examined halo 281, a lower-mass, isolated host with a quiet assembly history, to verify the ability of CWTHF to identify subhalos in a simpler case. This halo is included only in the statistical analysis for simplicity.

The zoom-in analysis retains the parameter settings \texttt{dens\_th}=4 and \texttt{n\_ref}=400 but increases \texttt{kw\_high} to 40 and raises \texttt{w\_resolution} to 85 to ensure sufficient sampling density in scale space. Using the same spatial grid, the isolated halo occupies only about 7\% of the grid points along one spatial dimension, which corresponds to approximately 0.035\% of the total grid volume. By processing only these occupied points, both memory usage and computational demand are reduced proportionally, enabling an extremely fine-grained analysis to be conducted at a higher resolution.

\begin{figure}[t!]
\centerline{\includegraphics[width=0.475\textwidth]{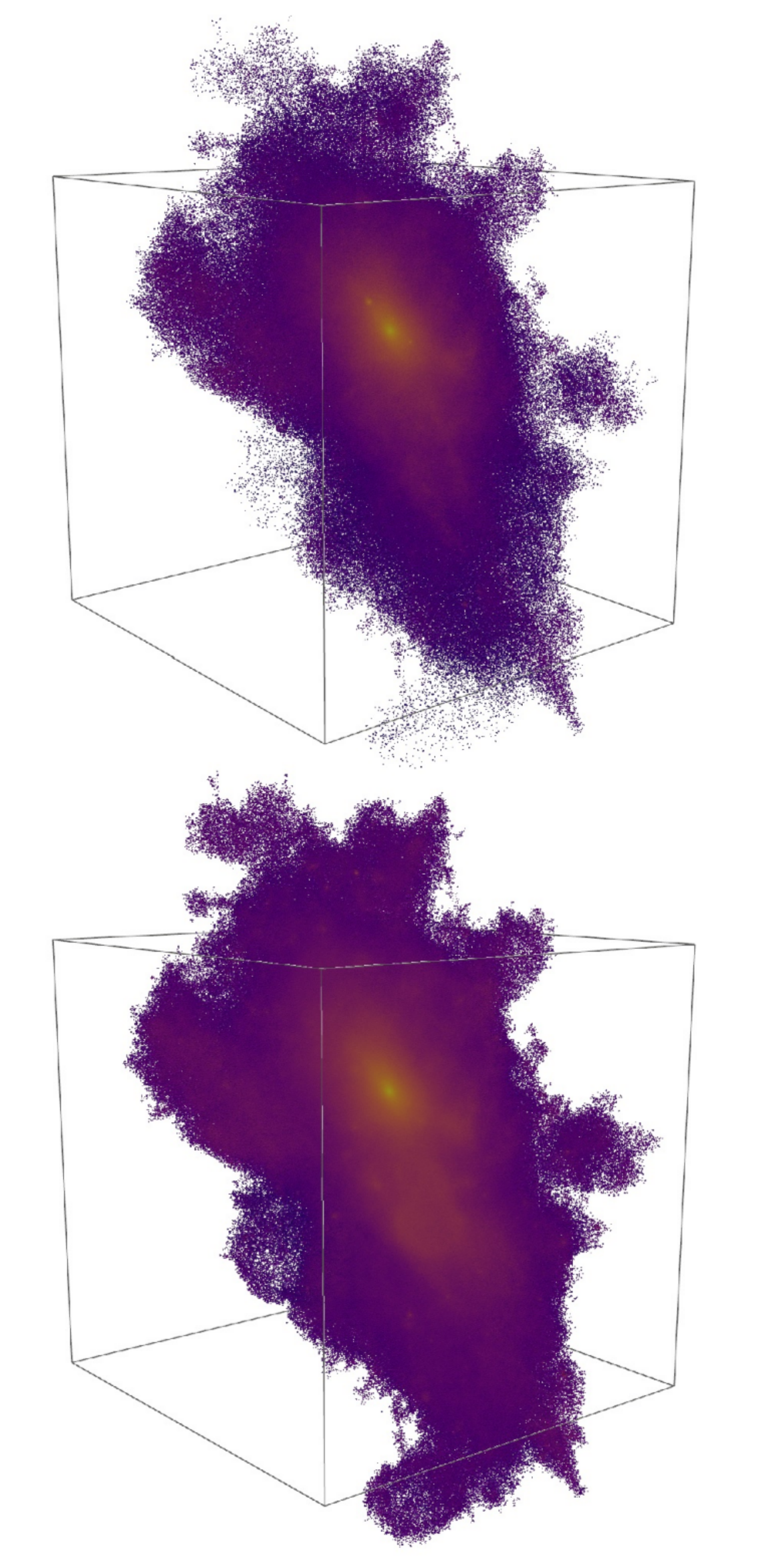}}
\caption{The central subhalos identified by CWTHF (top) and SUBFIND (bottom) from the halo in Figure~\ref{fig:largest_FOF}.}
\label{fig:FOF_c_subs}
\end{figure}

Within this refined analysis region, SUBFIND identifies 17,185 subhalos. Our zoom-in CWTHF procedure now identifies 14,895 subhalos, a number comparable to the SUBFIND result. Figure~\ref{fig:FOF_c_subs} shows the central subhalo as identified by the two different methods. Both exhibit an excellent unimodal structure. A notable difference, however, is the presence of 4–5 compact, luminous overdensities residing in the innermost region of the CWT-derived central subhalo. Halo 281 shows a similar pattern, except that it exhibits a perfect unimodal distribution with no additional overdensities, and has fewer particles and subhalos.

Undoubtedly, the CWT technique can successfully identify and separate subhalos from their background, provided an adequate search range is used in the scale dimension. Even in cases involving coupled central subhalos, these procedures were still capable of identifying and segmenting subhalos from the background halo. The only reason for their eventual mixture is that they failed to pass the subsequent self-boundness check. 

The unbinding procedure effectively addresses an inherent limitation of the wavelet technique: its difficulty in cleanly distinguishing bound structures from the surrounding background. While subtracting the background density can mitigate this issue \citep{Schwinn2018}, it cannot definitively determine which particles are genuinely bound to the subhalo. The unbinding step provides this critical discrimination.

The presence of residual subhalos demonstrates that the self-boundness check cannot reliably extract and verify all candidate particle groups against the background. The underlying limitation is straightforward. When iteratively removing unbound particles, the method assumes the initial estimates for the group's position and velocity are reasonably close to their true values. This assumption breaks down in dense core regions, where a substantial number of background particles become mixed into the CWT clump. If this contamination exceeds 50\%, the unbinding routine can no longer accurately capture the dynamical structure of the DM subhalo. In many cases, the target subhalo may be discarded entirely because its phase-space configuration differs from that of the dominant but spurious center, which comprises more than 50\% of the particles.

Lowering the particle survival threshold (the fraction below which a candidate group is considered unbound) does not effectively screen genuine subhalos from highly contaminated particle clumps. In fact, this adjustment tends to exacerbate the problem. A random patch of particles drawn from the background halo may survive the unbinding process partially intact. With a lower threshold, CWTHF is more likely to misclassify such background fluctuations as real subhalos, particularly because they can appear as coherent structures over larger wavelet scales.

Although this is a structural issue, where a subhalo is treated as a foreign object to be removed if it becomes too minor within the particle group, such loss occurs primarily when searching for the smallest subhalos nearest the center. Moreover, only a tiny fraction of subhalos survive after severe tidal stripping.

\begin{figure}[t]
\centerline{\includegraphics[width=0.475\textwidth]{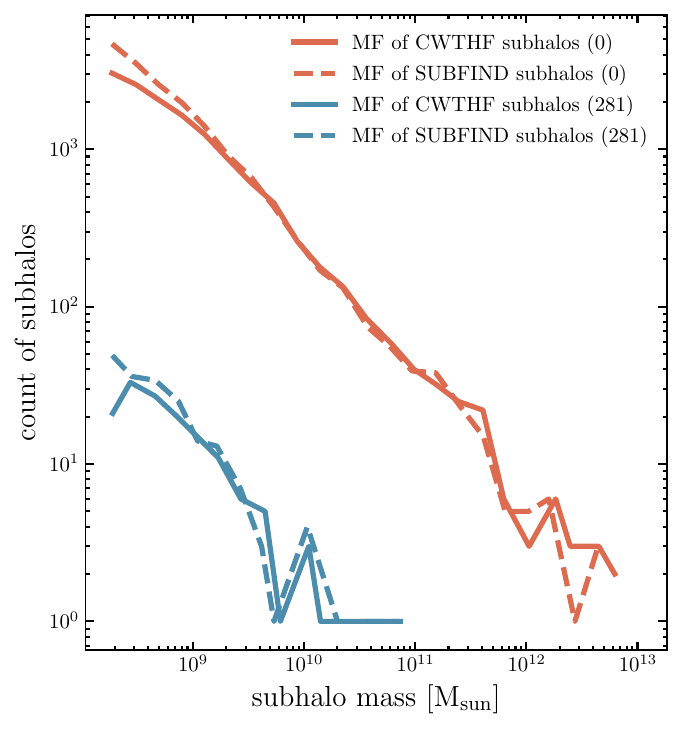}}
\caption{The SHMFs obtained by different methods within two halos. The red and blue lines represent halo 0 and halo 281, respectively, while the dashed and solid lines indicate the SUBFIND and CWTHF methods, respectively.}
\label{fig:HMF_sub}
\end{figure}

The decline in subhalo count is more clearly visible in the SHMFs, as shown in Figure~\ref{fig:HMF_sub}. For halo 0, the mass functions for large subhalos remain consistent between methods until the particle number falls below approximately 500, at which point the SHMF of CWT subhalos begins to bend downwards. As the particle number decreases further, the CWT catalog becomes increasingly incomplete, ultimately showing a deficit of about 40\% at the low-mass end. This trend is consistent with the contamination theory, i.e. smaller subhalos, with fewer self-bound particles, are more likely to be disrupted during the binding check by loosely associated background particles. Once again, the mass function of halo 281 shows almost the same pattern as halo 0.

The observed reduction in subhalo abundance is primarily attributed to the influence of background particle contamination. In the outer regions of the core, the gravitational effect of the central subhalo does not abruptly cease. For these low-mass subhalos, their limited particle content makes them highly susceptible to distortion by this residual background.

Once again, the highly consistent distribution of large subhalos confirms the robustness of CWTHF. However, it is also evident that CWTHF detects a lower population of subhalos compared to SUBFIND. Moreover, for the smallest subhalos, i.e., those with a half-mass radius of only a few $h^{-1}\rm kpc$, while their overall trend is quite similar, their spatial distributions do not fully coincide.

\begin{figure}[t!]
\centerline{\includegraphics[width=0.475\textwidth]{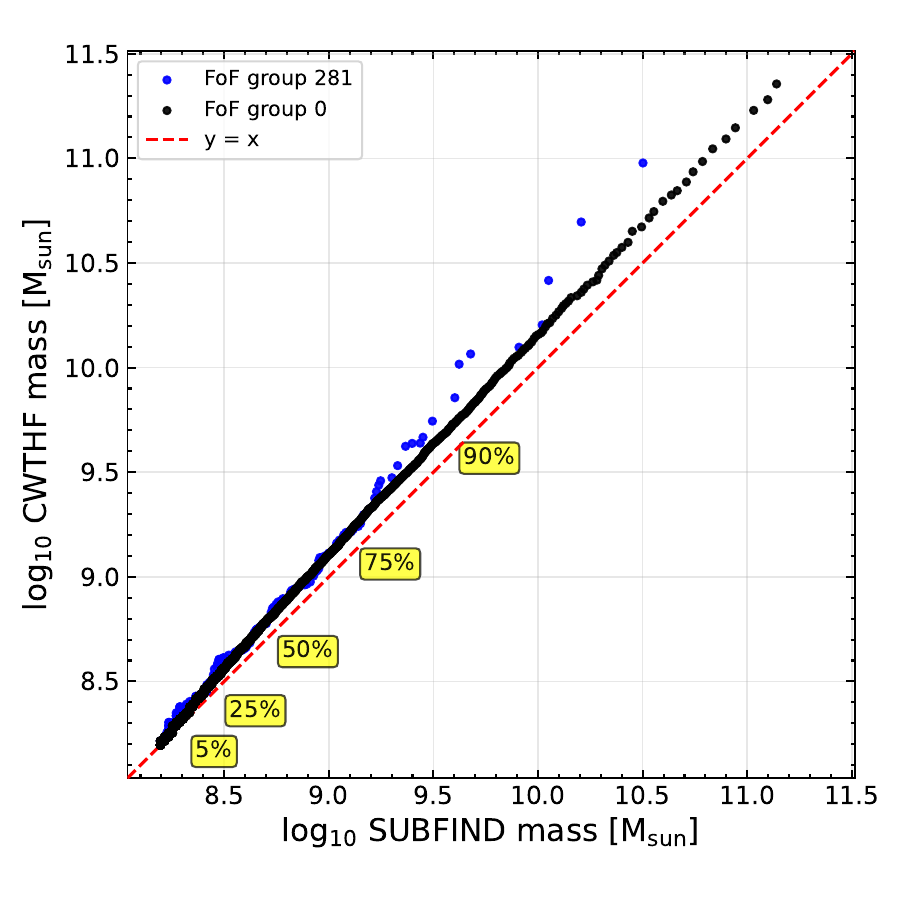}}
\caption{Quantile-quantile (Q-Q) plot comparing the subhalo masses identified by CWTHF and SUBFIND. Black and blue dots correspond to TNG halo 0 and halo 281, respectively. The red dashed line indicates the one-to-one relation ($y = x$). Yellow rectangles mark selected percentile values, which are approximately shared between the two halos.}
\label{fig:qq_plot}
\end{figure}

To further characterize the systematic difference in mass assignment between the two catalogs, we present a Q-Q plot in Figure~\ref{fig:qq_plot}, with data points drawn from the 0.5th to the 99.5th percentile. Although the two halos span different mass ranges and their distributions are not identical, their quantile positions are sufficiently similar that a single set of percentile markers is used for both, to avoid visual confusion.

Both sets of data points lie in close proximity to the $y = x$ reference line, indicating that the overall mass distributions recovered by CWTHF and SUBFIND are broadly consistent. Nevertheless, at any given percentile, the CWTHF subhalo masses are systematically higher than their SUBFIND counterparts. This offset is small at low percentiles but grows progressively toward the high-mass end, reaching approximately 0.5 and 1 order of magnitude for halo 0 and halo 281, respectively. We note that the central subhalo--i.e., the most massive member corresponding to the 100th percentile, which is excluded from Figure~\ref{fig:qq_plot}--is a notable exception, as it falls slightly below the $y = x$ line.

This systematic mass excess provides a direct explanation for the $\sim$30\% surplus in particle counts reported in Section~\ref{sec:full}. Note that this 0.5 dex offset at the high-mass end refers to the medium-to-large subhalos only, whereas the global ~30\% particle surplus is an average across all subhalos. This trend can likely be attributed to the spatial segmentation strategy inherent to CWTHF--the algorithm extends subhalo boundaries as far outward as possible, and for large subhalos, the coarser-scale resolution elements (grid cubes) span a larger physical volume, thereby incorporating more surrounding mass.

\begin{figure}[t]
\centerline{\includegraphics[width=0.475\textwidth]{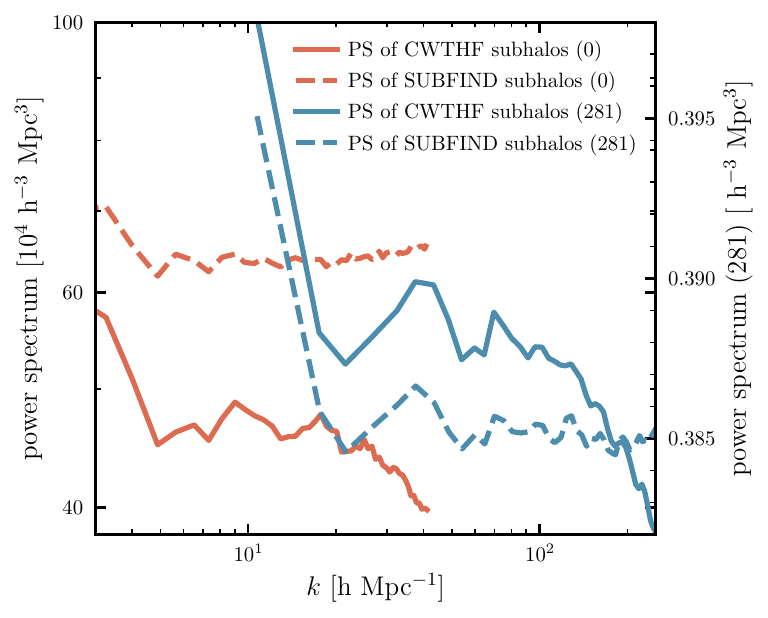}}
\caption{The SHPS obtained by different methods within two halos, with the line styles and color coding following the same scheme as in Figure~\ref{fig:HMF_sub}.}
\label{fig:PS_sub}
\end{figure}

Figure~\ref{fig:PS_sub} presents the SHPS for the two halos. The zoom-in SHPS is constructed by assigning all subhalos to a $128^3$ CIC grid confined within a box of length 5$\mpch$ and 1$\mpch$, respectively, matching the scale of each host halo. The SHPS of the CWT catalog is systematically lower in amplitude for halo 0, but higher for halo 281. As we included the central subhalo in those subhalo statistics, it is clear that the mass of the heavy central subhalo significantly affects the amplitude of those SHPS. Nevertheless, the good agreement for large subhalos ensures that the two SHPS curves follow similar evolutionary trends, and thus, after a suitable vertical offset, they exhibit nearly identical shapes.

\section{Discussion}
\label{sec:discussion}

Over the past two decades, a wide range of subhalo finders have been developed, which can be broadly categorized into configuration-space methods (e.g., HFOF, SUBFIND, AdaptaHOP), phase-space methods (e.g., ROCKSTAR), and history-space/tracking-based methods (e.g., HBT/HBT+). Setting aside the history-space information, the organization of subhalos in current finders is predominantly based on particles, while purely mesh-based approaches remain comparatively rare (with AHF representing an AMR-assisted hybrid). In this context, CWTHF extends the mesh-based route by segmenting the simulated space directly in the CWT grid, offering a methodologically orthogonal perspective for subhalo identification that is particularly well suited for cross-validation against particle-based finders.

Built upon the continuous wavelet transform, CWTHF provides a wavelet-space operational definition of halos and subhalos: structures correspond to local maxima of the CWT grid and the regions they dominate, supplemented by an unbinding step to ensure self-gravitational consistency. This definition not only admits a physical interpretation of what constitutes a halo within the CWT framework, but also yields a complete and executable identification pipeline.

CWTHF further exhibits substantial flexibility in scale selection. By specifying the range of $k_w$, the algorithm can target halos within a given physical scale range and isolate their well-resolved dominant subhalos. In addition, CWTHF can also operate as a standalone subhalo finder, performing subhalo decomposition on any given parent halo identified by other methods.

In terms of computational cost, the current implementation of CWTHF requires approximately $100\,\mathrm{GB}$ of memory and $\sim 30$ effective CPU hours per $10^{9}$ particles\footnote{Estimated from the run on the TNG100-1 dark matter sample ($\sim 6\times 10^{9}$ particles) on a 20-thread node within 24 hr of wall-clock time, accounting for the limited parallel efficiency of the self-boundness check and the residual serial portions of the code.}. For comparison, ROCKSTAR reports approximately $60\,\mathrm{GB}$ and $10$ CPU hours per $10^{9}$ particles in its reference implementation \citep{Behroozi2013}. The factor-of-a-few difference is largely attributable to language- and implementation-level constant factors rather than to the underlying algorithm: CWTHF maintains a strictly linear $\mathcal{O}(N)$ scaling.

This performance should be interpreted in light of the fact that CWTHF is currently Python-based, with critical kernels accelerated by Numba and Cython, which still leaves intrinsic performance limitations and significant potential for further optimization. In terms of memory usage, the CWT method performs only local calculations, allowing for potential domain decomposition strategies to reduce memory overhead. Furthermore, the single-halo test indicates that CWTHF could utilize a dynamic grid construction scheme to effectively achieve higher values of \texttt{kw\_high} within very massive halos. Additionally, we recognize a key limitation of the present algorithm: to fully realize the benefits of its cross-scale identification, a more efficient unbinding procedure, potentially operating directly on particles, is required to robustly mitigate contamination effects, especially for heavily stripped subhalos and the low-density outskirts of host halos.

Finally, a particularly promising future direction is to exploit the detailed CWT coefficients in an extended grid around each subhalo as a novel, spatially resolved signature for cross-snapshot matching. Through such feature extraction in the CWT space, CWTHF can be naturally extended along the time dimension toward the history-space regime, thereby enabling the construction of merger trees in a physically motivated way.

\section{Conclusions}
\label{sec:concl}

This work extends the CWT-based halo finder, i.e. the CWTHF algorithm, to identify substructures within dark matter halos. We use this algorithm to generate subhalo catalogs, which are then compared with the FOF-SUBFIND catalogs from the TNG simulation data. This comparison is conducted both for a full simulation snapshot and for an individual halo. The CWTHF code reuses the core computational framework of the original halo finder, thereby preserving its $\mathcal{O}(N)$ time complexity. The identification process begins by selecting parameters identical to those in our prior work, as their definitions remain unchanged. CWTHF proceeds by computing the CWT on the grid, locating the maxima of the transform, and segmenting the CWT grid based on these peaks. A key modification is that the locating and segmenting operations are now applied recursively within previously identified structures to find nested substructures. Each subhalo is defined by a cross-scale maximum, with its spatial boundary determined by aggregating all grid points where the CWT value decreases monotonically from that peak. Once a candidate passes the self-boundness check, its position is registered to prevent duplicate identification.

Following the construction of the CWT catalog, we conduct a series of validation tests for the updated CWTHF algorithm. We first compare the subhalo decomposition of the most massive CWT halo with its SUBFIND counterpart. While CWTHF identifies fewer subhalos within this host halo due to its more restricted search range, it successfully recovers all massive substructures in the hierarchy. For the other most massive central subhalos, their self-bound nature ensures that both their mass ranking and absolute mass remain stable across all test catalogs. We further compute their SHMFs and SHPS to quantify global population properties and to probe parameter dependencies. With parameter definitions identical to prior work, their influence on the derived subhalo distribution remains essentially unchanged.

To overcome the scale-resolution limit inherent in CWTHF, we perform a dedicated analysis on a single zoom-in FOF halo. The parameter setting of the spatial resolution remains unchanged, but the computation is restricted to only those grid cells associated with the selected halo. By increasing \texttt{kw\_high} to 40, CWTHF successfully resolves the majority of substructures embedded within the central subhalo. The ~30\% surplus in particle count relative to SUBFIND can largely be attributed to large subhalos. Besides that, a minor systematic artifact was observed: the contamination from background particles renders a few of the innermost and outermost smallest subhalos unbound, leading to their exclusion from the final catalog. These losses introduce a slight turnover in the SHMF and a corresponding downward shift in the SHPS, although the overall shape and evolutionary trends of both statistics remain unchanged.

Below we summarize the performance of the updated CWTHF and compare it with SUBFIND using both full-snapshot and isolated-halo tests.
\begin{enumerate}
  \item The CWT technique has been shown to effectively locate subhalo positions even within complex dynamical backgrounds. While the CWT by itself cannot distinguish subhalos from the background, this limitation is primarily addressed by implementing an additional unbinding procedure.

  \item For the largest central subhalos, their mass and rank exhibit remarkable stability across different catalogs, far exceeding the stability observed for the host halos themselves. The maximum relative difference in the subhalo mass is approximately 30\%, while the mean relative difference is only less than 10\% on average. Furthermore, the typical variation in rank order is less than four positions.

  \item We compute the SHMFs and SHPS to analyze their statistical properties. The SHMFs are largely consistent between the catalogs, except at the low-mass end of TNG100-1, where the lack of a high \texttt{kw\_high} value causes the function to bend downward. For the SHPS, the amplitude from CWTHF in TNG100-1 is higher across all scales. In TNG50-2, by contrast, the CWTHF power spectrum falls below that of SUBFIND as the scale increases.

  \item To verify the influence of the key parameters, we perform a parameter sensitivity test on the full simulation snapshot. The effects observed closely match those documented in our earlier study. Specifically, enhancing the resolution by increasing parameters such as \texttt{kw\_high}, \texttt{n\_ref}, and \texttt{w\_resolution} or lowering the density threshold \texttt{dens\_th}, leads to an upward bend in the low-mass end of the SHMFs and a systematic downward shift in the SHPS. Conversely, the opposite adjustments to these parameters produce the inverse effects.

  \item In a zoom-in analysis of the largest FOF halo in TNG100-1, the multiscale capability of our method successfully separates a clean, unimodal central subhalo from its host dark matter halo. This result demonstrates the viability of the CWT approach for locating and cleanly extracting substructures. The identified central subhalo closely matches the one found by SUBFIND, differing only by a few coupled subhalos in the innermost kernel region. Furthermore, the origin of the additional 30\% particles is illustrated by a Q-Q plot.

  \item The contamination from the central subhalo disrupts both the unbinding procedure for the innermost subhalos and the structure of smaller subhalos in the outer regions, leading to the characteristic downward bend in their SHMFs. Combined with the effects of parameter selection and the general absence of the smallest substructures, this results in a lower overall amplitude of the SHPS for CWTHF compared to SUBFIND.
\end{enumerate}

\section*{Acknowledgments}

The authors thank the anonymous referee for helpful comments and suggestions. We acknowledge the use of the IllustrisTNG simulation data. The authors used publicly available large language models to scout software libraries for standard functions, helping to avoid redundant work and streamline early development. All code implementations were independently verified and integrated by the authors. The source code for CWTHF is publicly available on Zenodo \citep{li_2025_17756729}\footnote{For future updated versions of the CWTHF, see \url{https://github.com/salty-fish-114514/CWTHF}.}

\software{NumPy \citep{vanderWalt2011,Harris2020}\footnote{\url{https://numpy.org/}}, SciPy \citep{Virtanen2020}\footnote{\url{https://scipy.org/}}, Matplotlib \citep{Hunter2007}\footnote{\url{https://matplotlib.org/}}, Jupyter Notebook\footnote{\url{https://jupyter.org/}}, Cython\footnote{\url{https://cython.org/}}}


\bibliography{Ref}{}
\bibliographystyle{aasjournal}
\end{CJK*}
\end{document}